\documentclass[a4paper]{article}

\usepackage{graphicx} 
\usepackage[top=20truemm,bottom=25truemm,left=25truemm,right=25truemm]{geometry}
\usepackage{amsmath, amssymb, bm, amsfonts, MnSymbol} 
\usepackage[dvipsnames]{xcolor}
\usepackage{cite}
\usepackage{authblk} 

\makeatletter 

\newcommand{\figref}[1]{Fig.~\ref{#1}}

\newcommand{\Eqref}[1]{Eq.~(\ref{#1})}

\makeatother 

\begin{document}

\title{{\bf Extension of Moving Particle Simulation including rotational degrees of freedom for dilute fiber suspension}}
\author[1]{Keigo Enomoto}
\author[1]{Takato Ishida}
\author[1]{Yuya Doi}
\author[1]{Takashi Uneyama}
\author[1]{Yuichi Masubuchi}
\affil[1]{Department of Materials Physics, Graduate School of Engineering, Nagoya University,
Furo-cho, Chikusa, Nagoya 464--8603, Japan}
\setcounter{Maxaffil}{0}
\renewcommand\Affilfont{\itshape}

\date{}

\maketitle

\begin{abstract}
  We develop a novel Moving Particle Simulation (MPS) method to accurately reproduce the motion of fibers floating in sheared liquids.
  In conventional MPS schemes, if a fiber suspended in a liquid is represented by a one-dimensional array of MPS particles, it is entirely aligned to the flow direction due to the lack of shear stress difference between fiber-liquid interfaces.
  To address this problem, we employ the micropolar fluid model to introduce rotational degrees of freedom into the MPS particles.
  The translational motion of liquid and solid particles and the rotation of solid particles are calculated with the explicit MPS algorithm.
  The fiber is modeled as an array of micropolar fluid particles bonded with stretching and bending potentials.
  The motion of a single rigid fiber is simulated in a three-dimensional shear flow generated between two moving solid walls.
  We show that the proposed method is capable of reproducing the fiber motion predicted by Jeffery's theory being different from the conventional MPS simulations.
\end{abstract}


\section{Introduction}
Fluid particle methods have been developed for simulations of multi-phase flows\cite{Liu2010, Gotoh2016, Gotoh2021}.
In the simulations of liquid-solid systems, the particles represent the included liquid and solid to possess local quantities such as velocity and pressure.
The motion of each particle is calculated according to interactions based on its discretized governing equation with neighboring particles within a certain distance.
The Moving Particle Simulation (MPS) method, developed by Koshizuka et al.\cite{Koshizuka1996}, is one of such methods along with Smoothed Particle Hydrodynamics (SPH)\cite{Gingold1977, Monaghan1994} and has been actively developed in recent years.
Following the original MPS, which employs a semi-implicit scheme\cite{Koshizuka1998}, high-precision schemes such as particle regularization schemes\cite{Xu2009} and improvements of the differential operator models\cite{Khayyer2011, Tamai2014} have been proposed.
Further developments for MPS have been being attempted for various issues including variable resolution schemes, theoretical error analysis, momentum conservation at interfaces, etc~\cite{Souto-Iglesias2013, Duan2019, Li2020}.

A possible direction for further improvement of MPS is the inclusion of rotational degrees of freedom for particles.
Such an aspect is necessary for fiber suspensions when the fiber is represented by a one-dimensional array of particles.
Let us consider a rotational motion of a fiber oriented in the flow direction under shear.
In conventional MPS schemes, this fiber is trapped in the fully aligned state due to the balance of particle interactions.
However, in reality, due to the difference of the shear stress between the interfaces in the shear gradient direction, the fiber exhibits periodic rotation as theoretically argued by Jeffery\cite{Jeffery1922}.
Although this problem has been known\cite{Meyer2020}, it has not been properly considered in most of the simulations for fiber suspensions with MPS\cite{Yashiro2011, Yashiro2012}.
In the conventional fluid particle method, viscous torque exerted by the fluid cannot be transferred to the motion of solid particles.

In this study, we propose a novel MPS method for fiber suspensions to reproduce the rotational motion of fibers in a correct manner.
To achieve this objective, we employ the micropolar fluid model to introduce an angular velocity field through the rotational degrees of freedom of the constituent particles\cite{Eringen1966a}.
To evaluate our method, we performed simulations of a single fiber suspended in the sheared Newtonian liquid.
The fiber is represented as an array of micropolar fluid particles connected with each other with stretching, bending, and torsional potentials.
We compare the fiber motion with Jeffery's theory\cite{Jeffery1922} to confirm that the fiber motion is correctly captured.
Details are shown below.


\section{Model and Simulation}

\subsection{Explicit MPS with rotational degrees of freedom}
In the MPS model, the dynamics of fluid velocity obey the continuum Navier-Stokes equation.
To incorporate the rotational degrees of freedom into the dynamics model, we employ the micropolar fluid model\cite{Eringen1966a} in which the angular velocity field is incorporated.
The conservation laws of linear and angular momentum are written as follows:
\begin{gather}
  \frac{D \bm{u}(\bm{r},t)}{D t}= -\frac{1}{\rho}\bm{\nabla} P(\bm{r},t)+ \nu \bm{\nabla}^2 \bm{u}(\bm{r},t) + \nu_r\bm{\nabla} \times \bm{\Upsilon}(\bm{r},t) + \bm{f}(\bm{r},t) ,
    \label{eq: trans} \\
  \mathcal{I}\frac{D \bm{\Omega}(\bm{r},t)}{D t}=  \bm{G}(\bm{r},t) - \nu_r\bm{\Upsilon}(\bm{r},t),
    \label{eq: rotate}
\end{gather}
where \( D/Dt \) is the time material derivative, \( \bm{r} \) is the position vector, \( t \) is time, \( \bm{u}(\bm{r},t) \) is the fluid velocity, \( \rho \) is the mass density,
\( P(\bm{r},t) \) is the pressure, \( \nu \) is the kinematic viscosity coefficient, \( \nu_r \) is the rotational kinematic coefficient, \( \bm{\Omega}(\bm{r},t) \) is the angular velocity field,
\( \bm{\Upsilon}(\bm{r},t) = 2\bm{\Omega}(\bm{r},t) - \bm{\nabla} \times \bm{u}(\bm{r},t) \),
\( \bm{f}(\bm{r},t) \) is the external volume force, \( \mathcal{I} \) is the micro-inertia coefficient, and \( \bm{G}(\bm{r},t) \) is the torque density due to the external field.
According to the second law of thermodynamics, \( \nu_r \) is a parameter properly chosen in the following range\cite{Souto-Iglesias2021}:
\begin{equation}
  0 \leq \nu_r \leq \left( 1 + \frac{2}{d} \right)\nu.
  \label{eq: nur}
\end{equation}
Here, \( d \) is the spatial dimension.
For a normal fluid without micropolar degrees of freedom, \( \bm{\Omega} \) is given as \( \bm{\Omega} =\left( \bm{\nabla} \times \bm{u}\right)/2 \) which guarantees that \Eqref{eq: trans} reduces the standard Navier-Stokes equation\cite{Souto-Iglesias2021}.
In this work, we simply set \( \bm{\Omega} = (\nabla \times \bm{u}) / 2 \) for liquid region.

In this study, we employ the explicit MPS (EMPS) method\cite{Oochi2010, Shakibaeinia2010} to discretize Eqs.~\eqref{eq: trans} and~\eqref{eq: rotate}.
The equations for the constituent particle \( i \) are as follows:
\begin{gather}
  \frac{d \bm{u}_i(t) }{d t}= -\frac{1}{\rho_i}\left\llangle \bm{\nabla}P \right\rrangle_i\!(t) + \frac{1}{\mathrm{Re}} \left\llangle \bm{\nabla}^2 \bm{u} \right\rrangle_i\!(t) +\frac{1}{\mathrm{Re}_r} \left\llangle \bm{\nabla} \times \bm{\Upsilon} \right\rrangle_i\!(t) + \bm{f}_i(t) ,
  \label{eq: trans-i} \\
  \frac{d \bm{\Omega}_i(t)}{d t}=  \alpha\bm{G}_i(t) - \frac{2\alpha}{\mathrm{Re}_r}\bm{\Upsilon}_i(t) ,
  \label{eq: rotate-i} \\
  \bm{\Upsilon}_i(t) = 2 \bm{\Omega}_i(t) - \left\llangle\bm{\nabla} \times \bm{u}\right\rrangle_i\!(t),
\end{gather}
where \( \llangle \rrangle \) indicates the quantity evaluated by the operator model in MPS at the position of particle \( i \) mentioned in the next paragraph.
The equations are non-dimensionalized using the following quantities: the fluid mass density \( \rho_0 \), the reference kinematic viscosity coefficient \( \nu_0 \), and the size of the fluid particle \( l_0 \).
\( \nu_0 \) is a reference value, and \( l_0 \) can be interpreted as the characteristic length scale of the discretized system (which may be interpreted as the grid size in the finite difference scheme).
These quantities define units of length, time, and energy, and the quantities discussed below are normalized according to these units.
\( \mathrm{Re}=\nu_0/\nu \) is the Reynolds number, \( \mathrm{Re}_r= \nu_0/\nu_r \) is the rotational Reynolds number, and \( \alpha \) is defined as \( \alpha=l_0^2/\mathcal{I} \).
As the case of the integration of the micropolar fluid model to the SPH model\cite{Souto-Iglesias2021}, translational and rotational velocities are mapped onto constituent (liquid and solid) particles.
In our model, solid particles are micropolar fluid particles, and their motion follows Eqs.~\eqref{eq: trans-i} and~\eqref{eq: rotate-i}.
The motion of liquid particles follows the standard Navier-Stokes equation plus the reaction force based on the third term on the right-hand side of \Eqref{eq: trans-i} exerted by the surrounding solid particles.

To calculate the physical quantities and their differentials at the position of particle \( i \), we need the weighting function.
We employ the following weighting function:
\begin{equation}
  w(r) =
  \begin{cases}
    l_c / r  \, -1  &(0 < r < l_c) \\
    0    &(r \geq l_c)
  \end{cases}.
  \label{eq:weight-function}
\end{equation}
Here, \( l_c \) is the cutoff radius.
The local density is evaluated by the local number density of the constituent particles defined as
\begin{equation}
  n_i = \sum_{j \neq i}w\left( \left | \bm{r}_j - \bm{r}_i \right |  \right).
  \label{eq:number-density}
\end{equation}
The differential operators in Eqs.~\eqref{eq: trans-i} and~\eqref{eq: rotate-i} are calculated by the following operator models:
\begin{align}
  \left\llangle \bm{\nabla} \psi \right\rrangle_i &= \frac{d}{n_0}\sum_{j\neq i}\left[\frac{\psi_i + \psi_j}{|\bm r_j - \bm r_i |^2}(\bm r_j - \bm r_i )w\left(|\bm r_j - \bm r_i |\right)\right] ,
  \label{eq: nabla-model} \\
  \left\llangle \bm{\nabla} \times \bm{b} \right\rrangle_i &= \frac{d}{n_0}\sum_{j\neq i}\left[\frac{\left( \bm{b}_j - \bm{b}_i \right) \times \left( \bm{r}_j - \bm{r}_i \right)}{|\bm r_j - \bm r_i |^2}(\bm r_j - \bm r_i )w\left(|\bm r_j - \bm r_i |\right)\right] ,
  \label{eq: rotation-model} \\
  \left\llangle \bm{\nabla}^2 \bm{b} \right\rrangle_i &= \frac{2d}{\lambda n_0}\sum_{j \neq i}\left[(\bm{b}_j - \bm{b}_i)w\left(|\bm r_j - \bm r_i |\right)\right] ,
  \label{eq: Laplacian-model} \\
  \lambda &= \frac{\sum_{j \neq i} {\left( \bm{r}_j - \bm{r}_i \right)}^2 w \left( \left | \bm{r}_j - \bm{r}_i \right |  \right)}
  {\sum_{j \neq i} w \left( \left | \bm{r}_j - \bm{r}_i \right |  \right) } .
  \label{eq: lambda}
\end{align}
Here, \( \psi_i \) and \( \bm{b}_i \) are scalar and vector variables on the particle \( i \), \( n_0 \) is the initial particle number density, and \( \lambda \) is the parameter defined by \Eqref{eq: lambda}\cite{Koshizuka1998}.
Note that in \Eqref{eq: nabla-model} we use \( \psi_i + \psi_j \) instead of \( \psi_j - \psi_i \), as proposed by Oochi et al.\cite{Oochi2010}, for better momentum conservation.

\subsection{Fiber model}
\begin{figure}[ht]
  \begin{center}
    \includegraphics[keepaspectratio,scale=0.35]{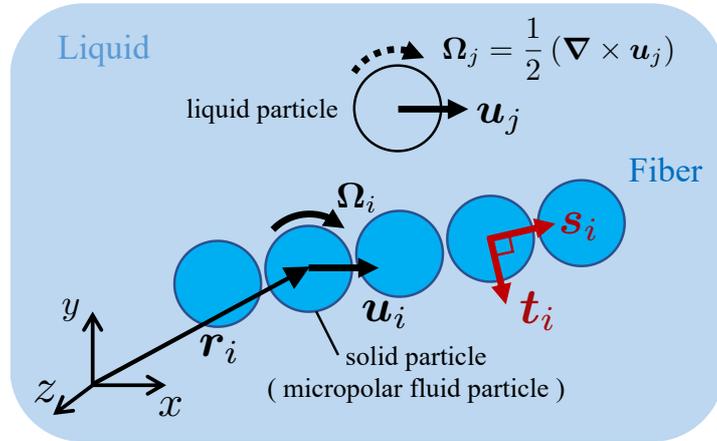}
  \caption{Schematic of our method. The fiber is composed of micropolar fluid particles which possess the velocity \( \bm{u}_i \) and angular velocity \( \bm{\Omega}_i \).
  The liquid particles are represented as a micropolar fluid particle with \( \bm{\Omega}_j = \left( \bm{\nabla} \times \bm{u}_j \right)/2 \).
  }
  \label{fig: model_concept}
\end{center}
\end{figure}
The fiber is modeled as an array of solid particles as shown in \figref{fig: model_concept}.
The solid particles are connected with stretching, bending, and torsional potential energies, in a similar manner proposed by Yamamoto and Matsuoka for the other simulation scheme\cite{Yamamoto1993}.
These potential forces should be a function of the bond vector of neighboring particles and the orientation of each solid particle\cite{Kuzkin2012}.
To describe the orientation of the solid particles, we introduce two directors \( \bm{s}_i \) and \( \bm{t}_i \) on each solid particle \( i \).
\( \bm{s}_i \) and \( \bm{t}_i \) are unit vectors for which directions are parallel and perpendicular to the bond vector, as shown in \figref{fig: model_concept}
The time derivative of directors is related to the angular velocity as follows:
\begin{equation}
  \frac{d \bm{s}_i(t)}{dt} = \left( \bm{1} - \bm{s}_i\bm{s}_i \right)\cdot \left( \bm{\Omega}_i \times \bm{s}_i \right) , \quad
  \frac{d \bm{t}_i(t)}{dt} = \left( \bm{1} - \bm{t}_i\bm{t}_i \right)\cdot \left( \bm{\Omega}_i \times \bm{t}_i \right).
  \label{eq:directors}
\end{equation}
Here, \( \bm{1} \) is the unit tensor.
The projection tensors \( (\bm{1} - \bm{s}_i\bm{s}_i) \) and \( (\bm{1} - \bm{t}_i\bm{t}_i) \) maintain \( \bm{s}_i \cdot \bm{t}_i = 0 \) within numerical errors.

The stretching potential \( U_s \), bending potential \( U_b \), and torsional potential \( U_t \) are defined as
\begin{align}
  U_s\left( \left\lbrace \bm{r}_i \right\rbrace \right) &= \sum_{\left\langle i,j \right\rangle}\frac{k_s}{2}{\left( \left | \bm{r}_j - \bm{r}_i \right | - 1 \right)}^2 ,
  \label{eq: pot_stretch} \\
  U_b\left( \left\lbrace \bm{r}_i \right\rbrace,\left\lbrace \bm{s}_i \right\rbrace \right) &= \sum_{\langle i,j \rangle} \left[  \frac{k_b}{2} {\left( \bm{s}_j - \bm{s}_i \right)}^2 - \frac{k_r}{2} \left\{ {\left( \bm{s}_i \cdot \frac{\bm{r}_j - \bm{r}_i}{|\bm{r}_j - \bm{r}_i|} \right)}^2 + {\left( \bm{s}_j \cdot \frac{\bm{r}_i - \bm{r}_j}{|\bm{r}_i - \bm{r}_j|} \right)}^2 \right\}\right] ,
  \label{eq: pot_bending} \\
  U_t\left( \left\lbrace \bm{t}_i \right\rbrace \right) &= \sum_{\langle i,j \rangle} \frac{k_t}{2} {\left( \bm{t}_j - \bm{t}_i \right)}^2 .
  \label{eq: pot_torsion}
\end{align}
Here, \( k_s, k_b, k_r, k_t \) are the spring constants and \( \left\langle i,j \right\rangle \) represents a pair of two adjacent solid particles.
The potential force \( \bm{f}_i \) and torque \( \bm{G}_i \) are calculated as
\begin{equation}
  \bm{f}_i = - \frac{\partial \left( U_s + U_b \right)}{\partial \bm{r}_i}, \quad
  \bm{G}_i = \bm{s}_i \times \left( - \frac{\partial U_b}{\partial \bm{s}_i} \right) + \bm{t}_i \times \left( - \frac{\partial U_t}{\partial \bm{t}_i} \right).
\end{equation}
According to \Eqref{eq: pot_stretch}, if \( k_s \) is large sufficiently, the fiber length \( L \) corresponds to the number of solid particles in the fiber.
Since the unit length of the system is the size of the fluid particle, the aspect ratio of the fiber \( r_p \) corresponds to \( L \).

\subsection{Numerical algorithms}

In the EMPS method, the fractional step algorithm is applied for time integration as in the original semi-implicit scheme for MPS\@.
Each integration step is divided into prediction and correction steps.
In the prediction step, predicted velocity \( \bm{u}_i^\ast \) is calculated by using terms other than the pressure gradient term in \Eqref{eq: trans-i}, and the angular velocity of the solid particles is also updated according to \Eqref{eq: rotate-i} as follows:
\begin{equation}
  \begin{aligned}
    \bm{u}_i^\ast &= \bm{u}^k_i + \Delta t \left[ \frac{1}{\mathrm{Re}} \left\llangle \bm{\nabla}^2 \bm{u} \right\rrangle_i^k + \frac{1}{\mathrm{Re}}_r \left\llangle \bm{\nabla}\times\bm{\Upsilon} \right\rrangle_i^k + \bm{f}_i^k \right] ,\\
    \bm{\Omega}_i^{k+1} &= \bm{\Omega}_i^{k} + \Delta t\alpha\left[ \bm{G}_i^k - \frac{2}{\mathrm{Re}_r}\bm{\Upsilon}_i^k \right] ,\\
    \bm{\Upsilon}_i^k &= 2\bm{\Omega}_i^k - \left\llangle \bm{\nabla} \times \bm{u} \right\rrangle_i^k .
  \end{aligned}
\end{equation}
Here, \( \Delta t \) is the step size, and the upper indexes \( k \) represent the step number: \( \bm{b}_i^k = \bm{b}_i (t = k\Delta t) \).
The predicted position \( \bm{r}_i^\ast \) and directors are updated as
\begin{equation}
  \begin{aligned}
    \bm{r}_i^\ast &= \bm{r}_i^k + \Delta t\bm{u}_i^\ast , \\
    \bm{s}_i^{k+1} &= \bm{s}_i^k + \Delta t\left( \bm{1} - \bm{s}_i^k\bm{s}_i^k \right)\cdot \left( \bm{\Omega}_i^{k+1} \times \bm{s}_i^k \right) ,\\
    \bm{t}_i^{\ast} &= \bm{t}_i^k + \Delta t\left( \bm{1} - \bm{t}_i^k\bm{t}_i^k \right)\cdot \left( \bm{\Omega}_i^{k+1} \times \bm{t}_i^k \right) ,
  \end{aligned}
\end{equation}
where \( \bm{t}_i^\ast \) is the predicted torsional director.
To maintain the relation \( \bm{s}_i \cdot \bm{t}_i = 0\), we adjust \( \bm{t} \) as follows:
\begin{equation}
  \bm{t}_i^{k+1} = (\bm{1} - \bm{s}_i^{k+1}\bm{s}_i^{k+1} ) \cdot \bm{t}_i^\ast.
\end{equation}

In the correction step, the velocity and position are calculated as
\begin{equation}
  \begin{aligned}
    \bm{u}_i^{k+1} &= \bm{u}_i^\ast - \frac{\Delta t}{\rho_i}\left\llangle \bm{\nabla}P \right\rrangle_i^{k+1} ,\\
    \bm{r}_i^{k+1} &= \bm{r}_i^\ast + \left( \bm{u}_i^{k+1} - \bm{u}_i^\ast \right)\Delta t .
  \end{aligned}
\end{equation}
In the EMPS, the pressure is calculated by the following explicit form\cite{Oochi2010}:
\begin{equation}
  P_i^{k+1} = \frac{\rho_i c_s}{n_0}\left( n_i^\ast - n_0 \right).
  \label{eq:pressure}
\end{equation}
Here, \( c_s \) is the sound speed, and \( n_i^\ast \) is the number density at \( \bm{r}^\ast \).
This \( c_s \) is optimized concerning reasonable incompressibility and numerical stability.

\subsection{Simulations}
We apply shear flows in the following boundary conditions.
Hereafter, we refer to flow, shear gradient, and vorticity directions as \( x \), \( y \), and \( z \) directions.
We employ periodic boundary conditions for \( x \) and \( z \) directions, whereas we place solid walls at \( y=0 \) and \( h \) perpendicular to the \( y \) direction.
These walls consist of three layers of liquid particles, which are fixed on a squared lattice.
Following the earlier study\cite{Yashiro2012}, we move the walls toward the \( x \) direction with the speed of \( u_{\mathrm{wall}} = \pm \dot{\gamma}h/2 \), where \( \dot{\gamma} \) is the apparent shear rate.
We have confirmed that the actual shear rate is equal to \( \dot{\gamma} \) and uniform throughout the system within a numerical error, in simulations without fibers, as shown in~\ref{sec: fluid_flow}.

\begin{figure}[ht]
  \begin{center}
    \includegraphics[keepaspectratio,scale=0.3]{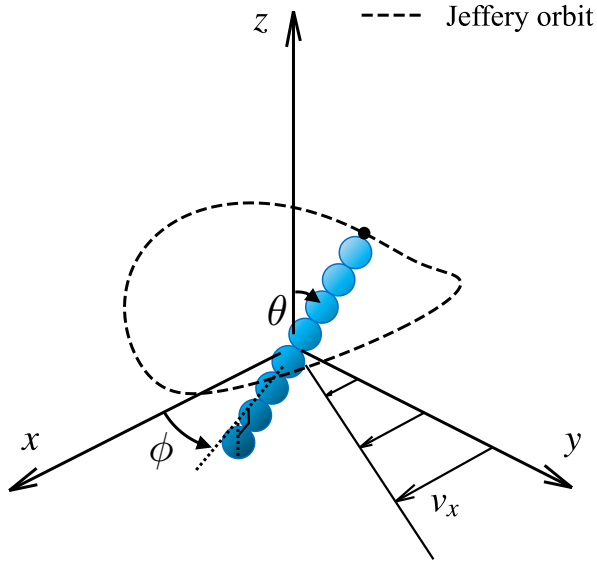}
  \caption{
    Schematic of a fiber (an array of blue particles) at orientation angles \( \phi \) and \( \theta \) in a shear flow.
    The dashed curve shows the orbit of the head of the fiber (Jeffery orbit).
  }
  \label{fig: diagram_Jeffery_orbit}
  \end{center}
\end{figure}
Simulations of a single fiber in a simple shear flow were carried out, and the rotational motion of the fiber was observed.
To describe the fiber motion, we use the orientation angles \( \phi \) and \( \theta \) as shown in \figref{fig: diagram_Jeffery_orbit}.
The number of MPS particles was \( N = 64000 \) in total including those for walls and the fiber.
The simulation box dimension was \( 40\times40\times40 \) in \( x \)-\( y \)-\( z \) directions, respectively, and the distance between the walls was \( 35 \).
The kinematic viscosity coefficient \( \nu \) and the strain rate \( \dot\gamma \) were chosen so that the fiber-based Reynolds number was \( \mathrm{Re}_f = L^2\dot\gamma/\nu = 0.1 \) to realize a viscous dominant condition.
The sound speed of the fluid \( c_s \) was set so that the Mach number became \( \mathrm{Ma} = 0.5h\dot\gamma/c_s = 0.03 \).
The numerical step size \( \Delta t \) was chosen to \( 0.01 \) according to the Courant condition, the viscous constraint, and the relation to the spring constant.
Other model parameters were set as \( l_c = 3.1 \), \( \nu_r = 1.5 \nu \), \( \mathcal{I} = 0.8 \) unless otherwise noted.
The mass density of the solid particles is the same as that of liquid particles.
The aspect ratio of the fiber \( r_p \) was varied in the range from 2 to 20.
The spring constants were chosen at \( k_s = 1000 \) and \( k_b = k_t = k_r = 200 \).
These values realized a rigid fiber, for which the effect of fiber deformation is negligible in the result as shown later.
We performed the simulations with a house-made code.

In the initial condition, we placed the fiber at the center of the simulation box to overlap the center of mass of the fiber and the box.
The initial fiber orientation angle to \( x \)-direction, \( \phi_0 \), was fixed at \( \pi/2 \), whereas the initial angle to \( z \)-direction, \( \theta_0 \), was chosen at \( \pi/6 \), \( \pi/3 \), or \( \pi/2 \).
Surrounding liquid particles were randomly arranged by the particle packing algorithms proposed by Colagrossi et al.\ \cite{Colagrossi2012}.

\section{Results and Discussion} \label{sec: results}

\begin{figure}[ht]
  \begin{center}
    \includegraphics[keepaspectratio,scale=0.55]{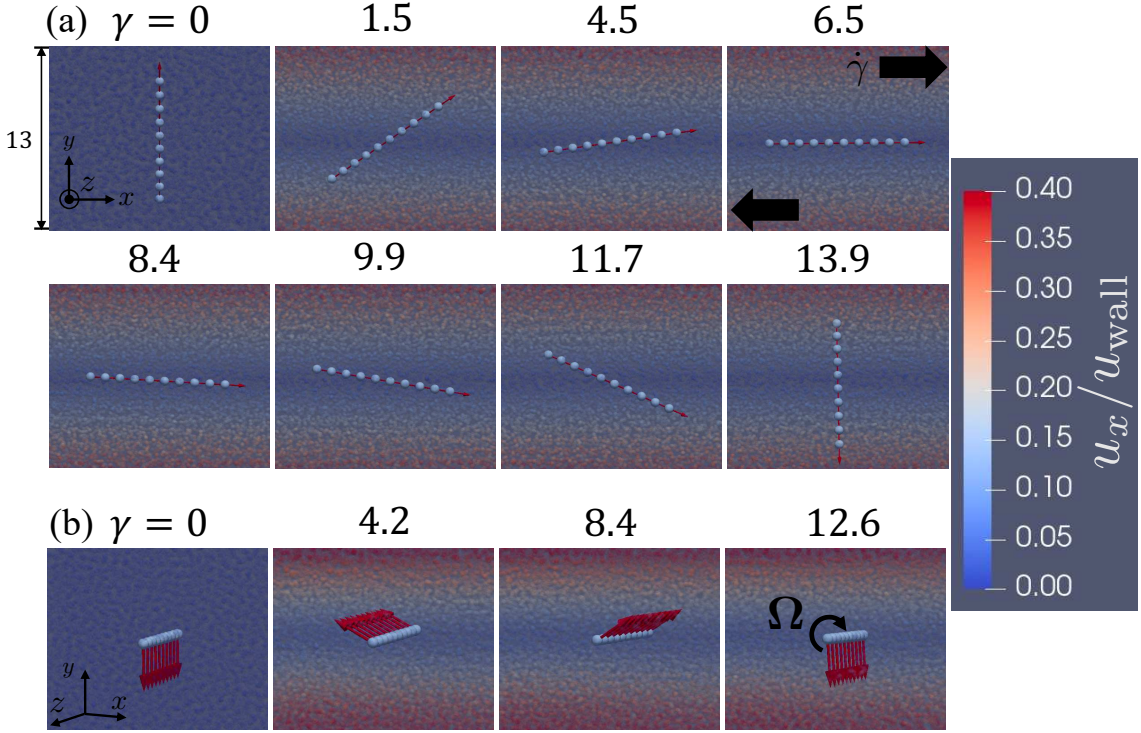}
  \caption{
    Typical snapshots of a fiber with \( r_p = 10 \).
    Light blue spheres represent solid particles that compose the fiber, and red arrows show directors.
    Background colors correspond to the velocity of fluid particles.
    (a) The case of \( \phi_0 = \pi / 2 \) and \( \theta_0 = \pi /3 \).
    The red arrows show \( \bm{s}_i \).
    (b) The case of \( \phi_0 = \pi / 2 \) and \( \theta_0 = 0 \).
    The red arrows show \( \bm{t}_i \).
  }
  \label{fig: snapshot}
  \end{center}
\end{figure}
Typical snapshots of a single rigid fiber in a shear flow with \( \theta_0 = \pi/3 \) are shown in \figref{fig: snapshot}~(a).
These figures clearly demonstrate that the fiber rotates as expected, even after it experiences the configuration aligned to the flow direction.
Snapshots of another fiber aligned to the vorticity direction (\( \theta_0 = 0 \)) are also shown in \figref{fig: snapshot}~(b).
The fiber exhibits the rolling motion around the vorticity axis induced by the flow velocity difference between shear planes above and below the fiber.
This behavior is known as the log-rolling motion\cite{Einarsson2015}.
In principle, we cannot reproduce this log-rolling motion of the fiber using MPS without introducing rotational degrees of freedom.


\begin{figure}[ht]
  \begin{center}
    \includegraphics[keepaspectratio,scale=0.5]{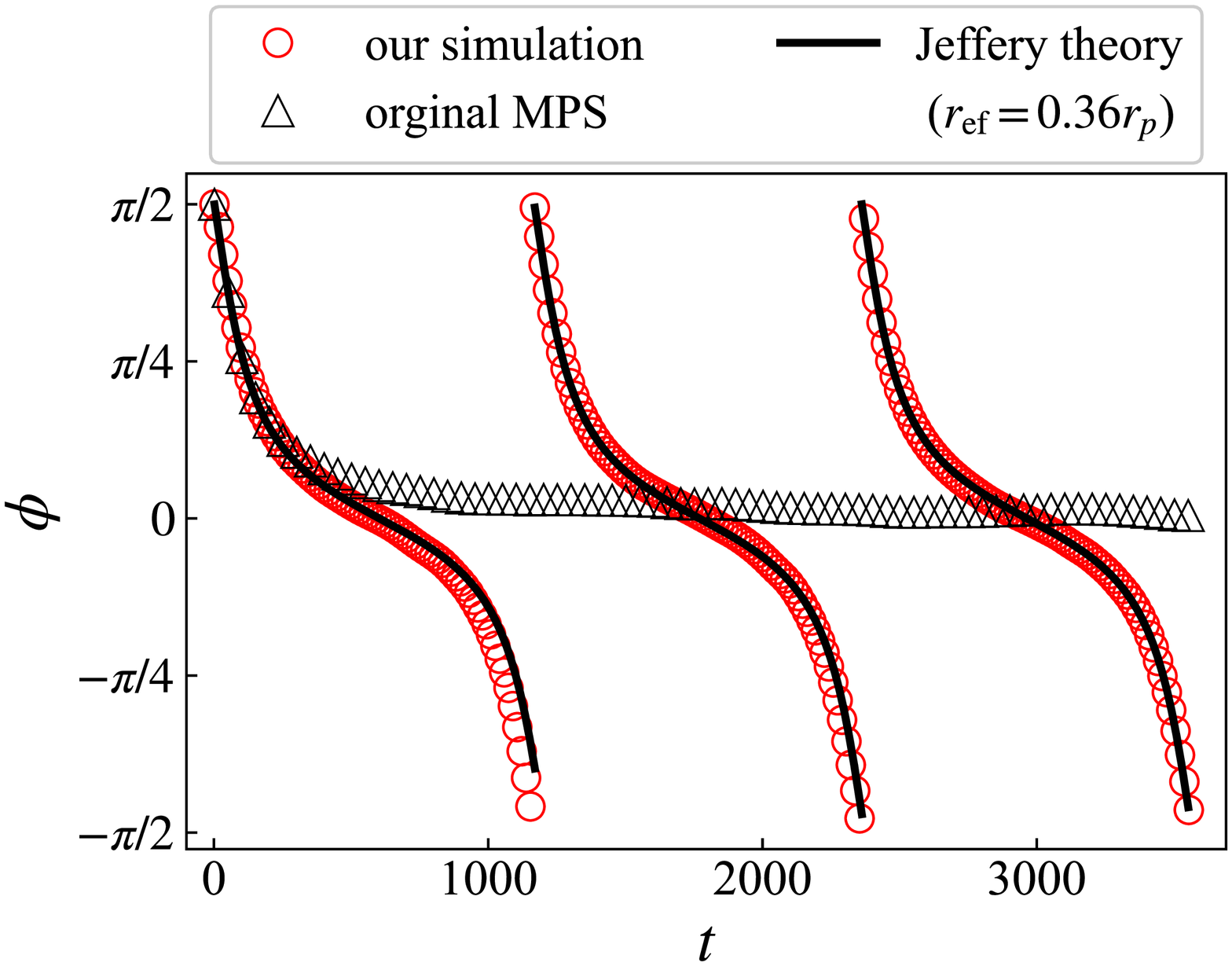}
    \caption{
      Time evolution of \( \phi \) by our model (circle) and the MPS without rotational degrees of freedom (triangle) in dilute regime. (\( r_p = 10, \phi_0 = \pi/2, \theta_0 = \pi/2 \))
      Solid curves represent Jeffery's theory with \( r_{\mathrm{ef}} = 0.36r_p \).
      }
    \label{fig: phi-t}
  \end{center}
\end{figure}
To analyze rotational behavior in the vorticity plane quantitatively, we show the time evolution of the rotation angle \( \phi \) in \figref{fig: phi-t}.
We observe that the fiber rotates and approaches to \( \phi = 0 \) in the MPS without rotational degrees of freedom.
This is not consistent with Jeffery's theory which predicts the periodic motion.
In contrast, in our model, we observe the clear periodic motion.
This fact demonstrates the importance of the rotational degrees of freedom integrated into our model.
We compare the time evolution of \( \phi \) with Jeffery's theory.
According to Jeffery's theory, the periodic orbit depends on the aspect ratio of the fiber.
The aspect ratio can be defined as the ratio of two axes of hydrodynamically equivalent ellipsoid for the fiber\cite{Bretherton1962}.
Here, one may argue that the fiber in our simulation model is not a rigid body and thus the aspect ratio is not well-defined.
We found that with the employed simulation parameters, the fiber almost keeps its length and shape under the flow, and thus it can be approximately treated as a rigid body.
We use the effective aspect ratio \( r_{\mathrm{ef}} = 0.36 r_p \) to achieve the best agreement between our model and Jeffery's theory.


\begin{figure}[ht]
  \begin{center}
    \includegraphics[keepaspectratio,scale=0.35]{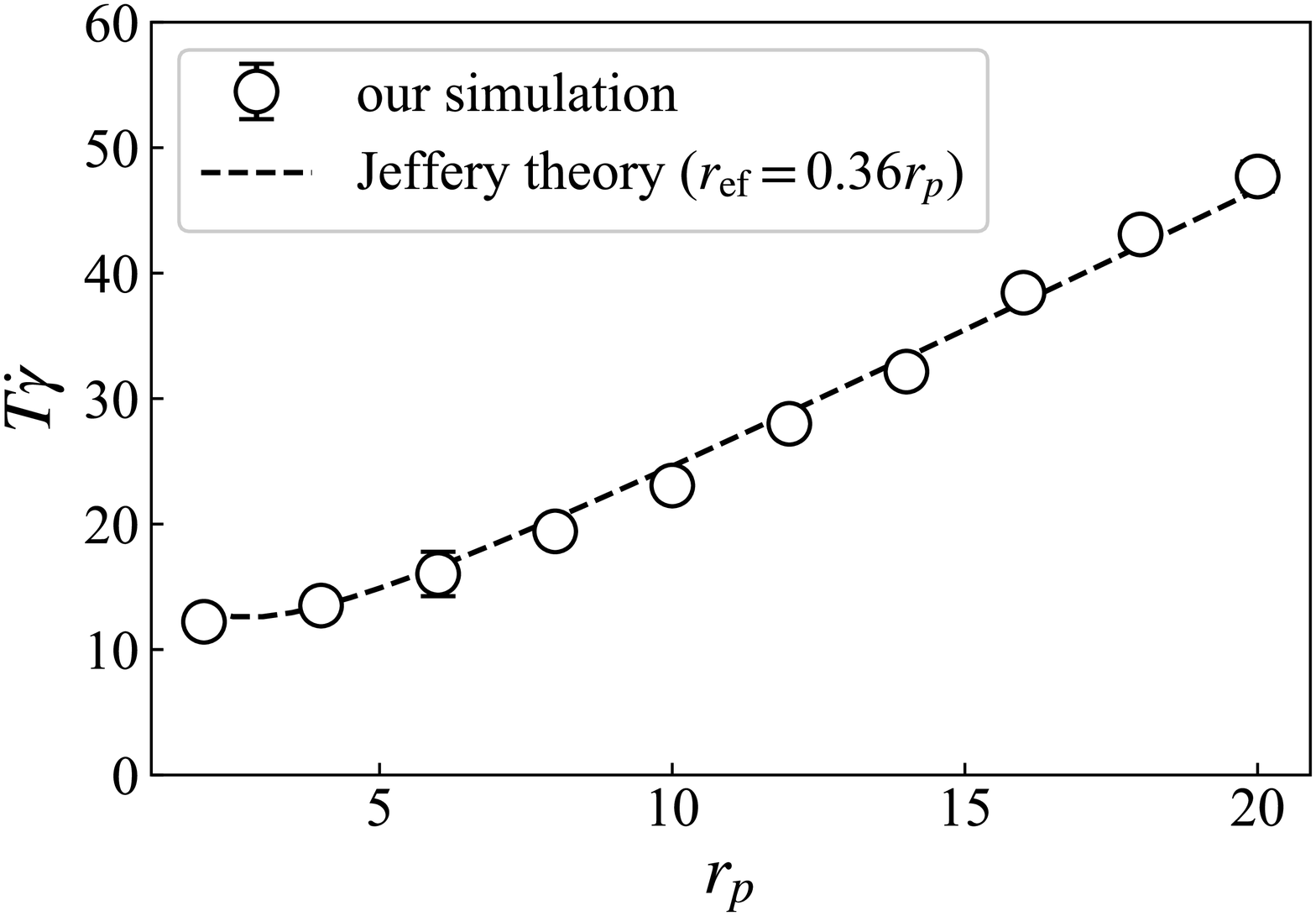}
    \caption{
      The aspect ratio dependence of the rotation period.
      Symbols show our simulation data and the dashed curve shows the prediction by Jeffery's theory (Eq. (23)) with \( r_{\mathrm{ef}} = 0.36 r_p \).
    }
    \label{fig: T-rp}
  \end{center}
\end{figure}
We performed the simulation with various aspect ratios to examine its effect on the rotation period.
The result is shown in \figref{fig: T-rp}.
According to Jeffery\cite{Jeffery1922}, the rotation period of the fiber \( T \) is described as
\begin{equation}
  T = \frac{2\pi}{\dot\gamma}\left( r_{\mathrm{ef}} + \frac{1}{r_{\mathrm{ef}}} \right) .
  \label{eq: T_Jeffery}
\end{equation}
As mentioned above, Jeffery's theory with the effective aspect ratio \( r_{\text{ef}} = 0.36 r_p \) agrees with our simulation data for \( r_p = 10 \).
We use the same relation for other \( r_p \) values.
As observed in Fig. 5, our simulation data agree well with Jeffery's theory with \( r_{\text{ef}} = 0.36 r_p \) within the examined \( r_p \) range.

The ratio \( r_{\mathrm{ef}} / r_p = 0.36 \) is not close to unity.
Here, we briefly discuss the validity of this value.
A typical value in experiments is \( r_{\text{ef}} / r_p = 0.7 \)\cite{Trevelyan1951}.
This value is larger than ours.
If we calculate the ratio of these two values, we have \( 0.7 / 0.36 \approx 1.9 \).
One interpretation of this result is that the fiber width in our model is twice larger than the expected value.
Intuitively, we expect that the motion of fluid particles around the fiber is somewhat synchronized and increases the effective width of the fiber.
Note that this ratio \( r_{\mathrm{ef}} / r_p \) depends on \( \nu_r \) as shown in~\ref{sec: nur_ref}.


\begin{figure}[ht]
  \begin{center}
    \includegraphics[keepaspectratio,scale=0.4]{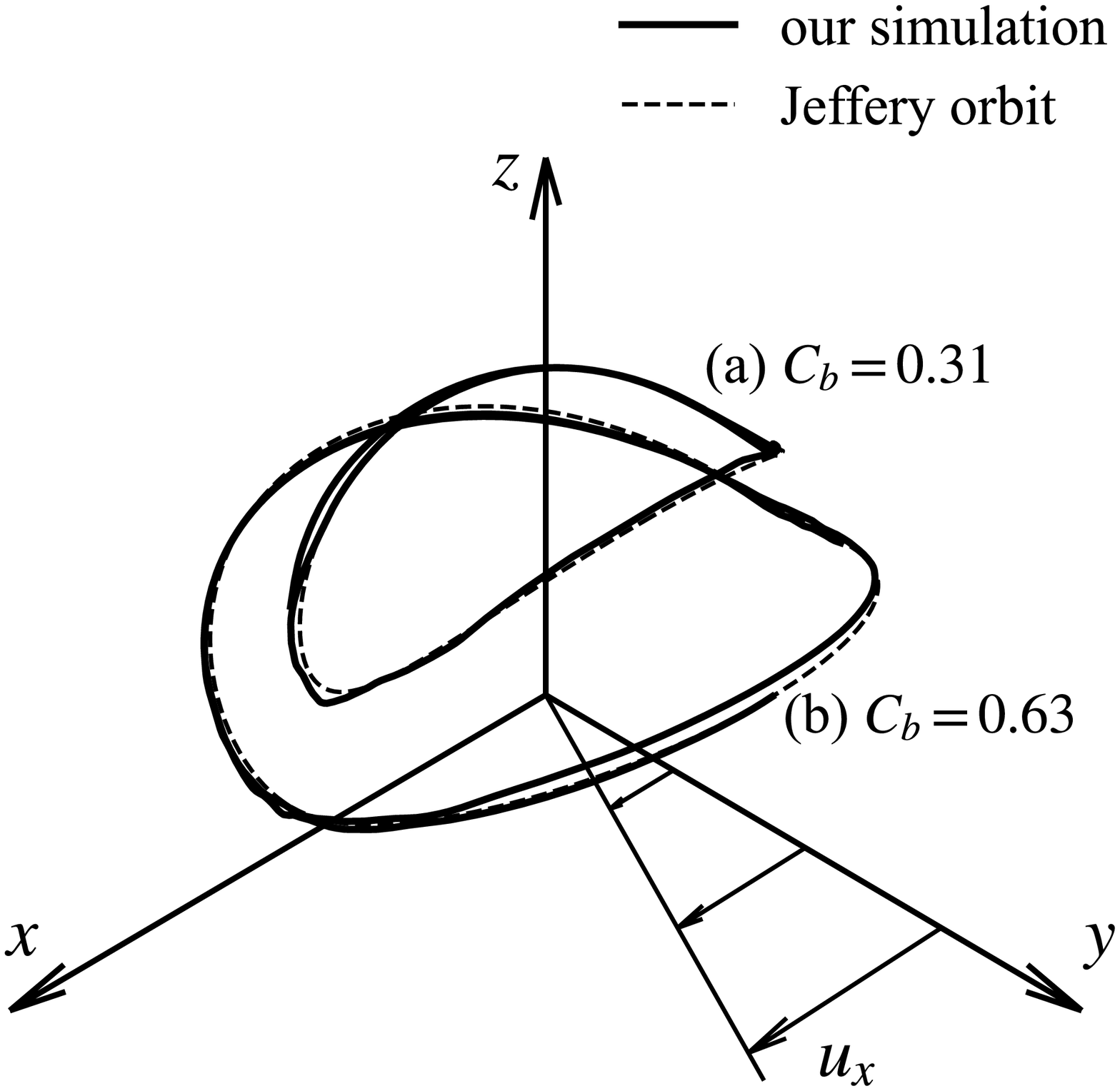}
    \caption{
      Rotation orbits of a single fiber.
      Solid curves show our simulation results of (a) \( \theta_0 = \pi/6 \) for \( 12 \leq \gamma \leq 50 \) and (b) \( \theta_0 = \pi/3 \) for \( 0 \leq \gamma \leq 50 \).
      Other parameters are the same as \figref{fig: snapshot} (a).
      Dashed curves are the Jeffery orbits with \( r_{\mathrm{ef}} = 0.36r_p \).
    }
    \label{fig: orbits}
  \end{center}
\end{figure}
We further examine pivoting motion of fibers that tilt from the vorticity plane.
\figref{fig: orbits} shows typical rotation orbits of the head of fibers for (a) \( \theta_0 = \pi/3 \) and (b) \( \theta_0 = \pi/6 \) with \( \mathrm{Re} = 0.01 \).
These orbits are characterized by \( C_b \) defined as
\begin{gather}
  C_b = |C_J|/\left( 1+|C_J| \right),
  \label{eq: Cb} \\
  C_J = \frac{1}{r_{\mathrm{ef}}}\tan\theta_0 {\left( r_{\mathrm{ef}}^2\sin^2\phi_0 + \cos^2\phi_0 \right)}^{\frac{1}{2}},
  \label{eq: CJ}
\end{gather}
where \( C_J \) is the orbit constant determined only by the initial configuration of the fiber \( \phi_0 \) and \( \theta_0 \).
The examined cases correspond to \( C_b = 0.31 \) and \( 0.63 \), respectively.
Although there are small fluctuations due to discretization errors, the fibers reasonably follow closed trajectories, which are consistent with the Jeffery orbits.

To be fair, we note that the fiber in our method eventually falls out of the Jeffery orbit if we continue the simulation for a long time.
Such behavior would be attributed to the properties of the Jeffery orbit and our model.
The Jeffery orbit is not stable against a perturbation\cite{Saffman1956}.
If a fiber motion or flow field is slightly perturbed, the orbit moves to others.
In our model, due to the discretization by using particles, both the fiber motion and flow field contain fluctuations.
These fluctuations drive the orbit away from the original Jeffery orbit.
We also note that the solid walls in our system and fluid inertia may probably play some roles.
Nevertheless, as shown in \figref{fig: orbits}, our scheme reasonably reproduces the Jeffery orbit in a similar manner to the other numerical studies\cite{Fan1998, Yamamoto1994, Skjetne1997a}.
Since our method is capable of reproducing the motion of single fibers in the dilute regime, extensions to the concentrated regime or real industrial application would be readily achievable.


\section{Conclusion}

We have developed a new MPS method to accurately reproduce fiber motion in shear flows.
We employ the micropolar fluid model to introduce rotational degrees of freedom into constituent particles.
To validate our method, we simulated the single fiber motion suspended in the sheared liquid.
The fiber is represented by a single array of micropolar fluid particles bonded with stretching, bending, and torsional potentials.
We demonstrated that the simulated rotation period and rotation orbits of the fiber are in good agreement with Jeffery's theory given that the effective aspect ratio is tuned as a fitting parameter of the theory.

As an application of the proposed method, we are conducting simulations for dense fiber suspensions since fiber rotation possibly plays some roles as argued by Lindstr{\"{o}}m and Uesaka\cite{Lindstrom2009}.
The proposed method is also capable of representing solids of arbitrary shape such as plate-shaped particles\cite{Sasayama2022}, not just fibers.
We are aware that the micropolar fluid model can be implemented to other fluid particle methods such as SPH\@.
Studies toward such directions are ongoing and the results will be reported elsewhere.

\section*{Acknowledgement}
The authors would like to express their gratitude to Dr.~Satoru Yamamoto at Center for Polymer Interface and Molecular Adhesion Science,
Kyushu University for helpful discussions.


\appendix
\def\thesection{Appendix \Alph{section}} 

\makeatletter
\def\theequation{\Alph{section}.\arabic{equation}}
\@addtoreset{equation}{section}
\makeatother

\section{Calculation of a simple shear flow using EMPS} \label{sec: fluid_flow}
\begin{figure}[ht]
  \begin{center}
    \includegraphics[keepaspectratio,scale=0.56]{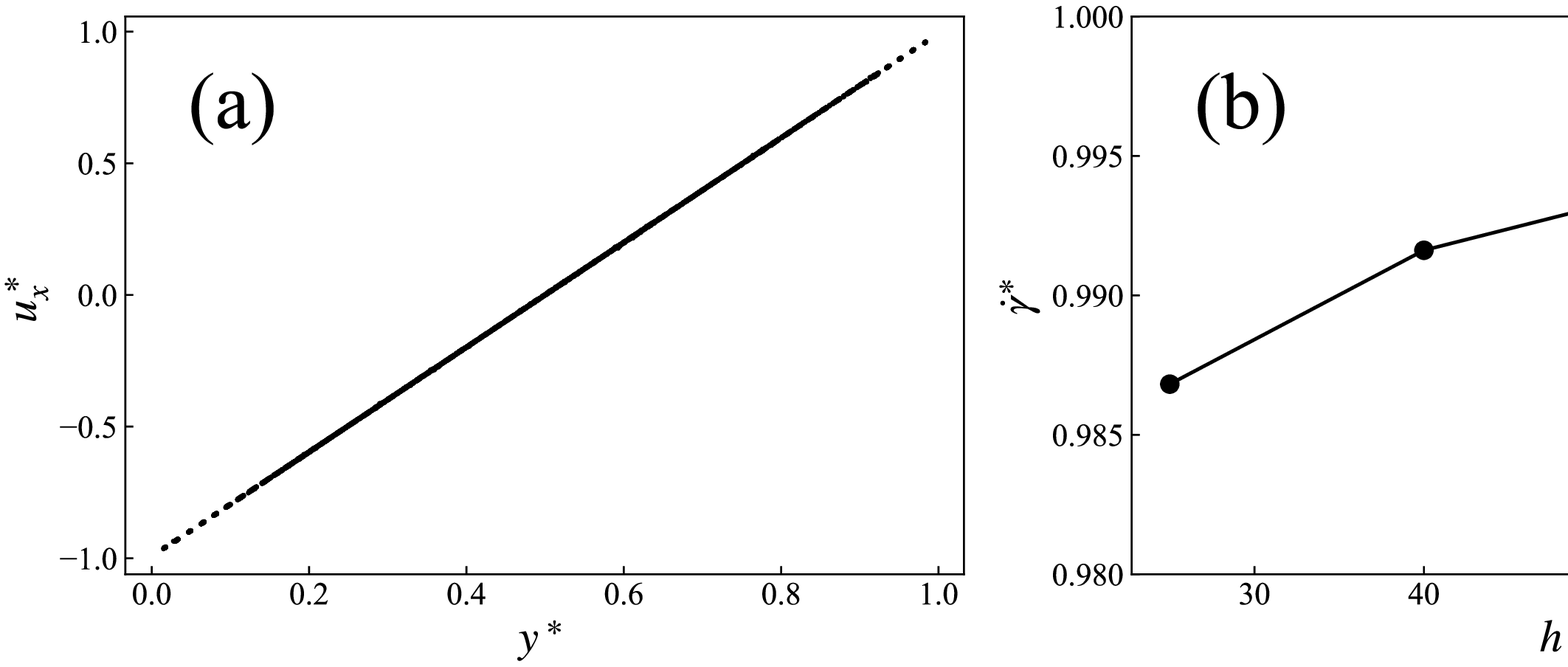}
    \caption{
      (a) Flow velocity profile generated by moving walls in our numerical method for a fluid without a fiber. The gap length is \( 55 \).
      (b) The gap length dependence of the average shear rate.
    }
    \label{fig: vel_profile}
    \end{center}
  \end{figure}

We have conducted EMPS simulations without solid particles to test the method and the code.
The system settings are the same as simulations in Sec. 3 except for the gap size \( h \) and the absence of a fiber.
An example of the steady-state flow profile of a shear flow is shown in \figref{fig: vel_profile} (a).
Here, \( u^\ast_x = u_x / u_{\mathrm{wall}}\) is the normalized fluid velocity in the flow direction (\( x \)-- direction) where the wall velocity \( u_{\mathrm{wall}} \), and \( y^\ast = y/h \) is the normalized distance from the moving wall.
The Reynolds number of the flow is \( \mathrm{Re}_h = h u_{\mathrm{wall}} / \nu = 1.5 \) for \( h = 55 \).
The result is in good agreement with the analytical solution \( u^{\ast}_x = 2 (y/h - 0.5)\).
The gap size dependence of the shear rate is shown in \figref{fig: vel_profile} (b).
Here, \( \dot\gamma^\ast \) is the average slope of the velocity profile divided by the shear rate expected from the wall velocity.
This result shows that the numerical error of the shear rate is less than 1\%  for \( h > 40 \).
These results are consistent with the earlier study\cite{Yashiro2012}.


\section{\( \nu_r \) dependence of the effective aspect ratio} \label{sec: nur_ref}

\begin{figure}[ht]
  \begin{center}
    \includegraphics[keepaspectratio,scale=0.35]{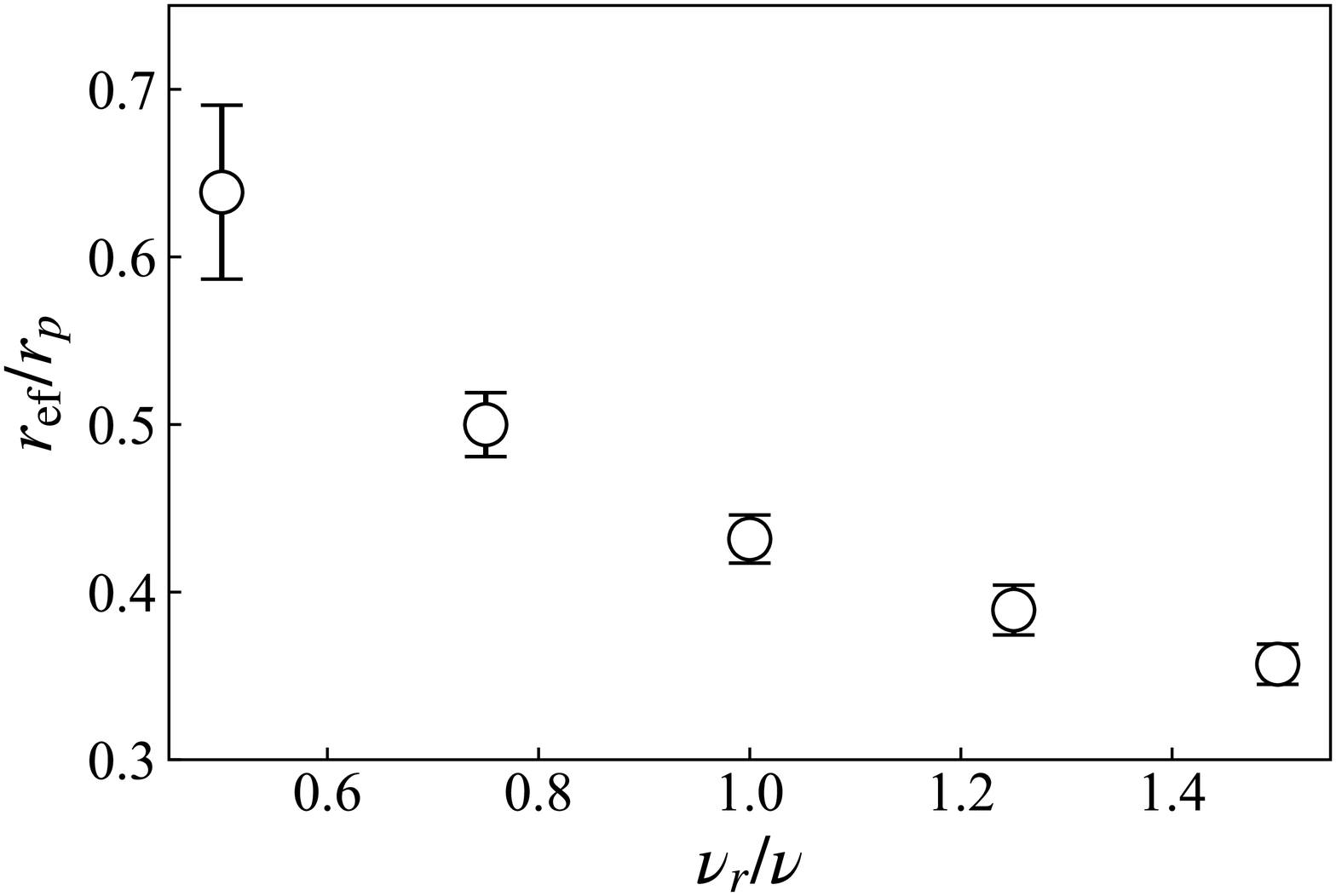}
  \caption{
    The rotational kinematic viscosity dependence of the effective aspect ratio in our model.
    \(r_{\mathrm{ef}} \) and \( \nu_{r} \) are normalized by \( r_p \) and \( \nu \), respectively.
  }
  \label{fig: ref-nur}
  \end{center}
\end{figure}

As mentinoed in Sec.~\ref{sec: results}, the ratio \( r_{\mathrm{ef}}/r_p \) depends on \( \nu_r \).
The results are shown in \figref{fig: ref-nur}.
As \( \nu_r \) increases, \( r_{\mathrm{ef}}/r_p \) monotonically decreases.
Thus, one may optimize \( \nu_r \) to have \( r_{\mathrm{ef}}/r_p \) that is consistent with a specific experimental system.
To be fair, we note that the conditions \( \nu_r < 0.5 \) are not suitable to our numerical method, and we cannot attain \( r_{\mathrm{ef}}/r_p \) value larger than 0.7, because the torque exerted to solid particles becomes comparable to the discretization error.


\bibliographystyle{unsrt}
\bibliography{Main_abbre}

\begin{thebibliography}{10}

\bibitem{Liu2010}
M.~B. Liu and G.~R. Liu.
\newblock {\em Arch. Comput. Methods Eng.}, 17(1):25--76, mar 2010.

\bibitem{Gotoh2016}
Hitoshi Gotoh and Abbas Khayyer.
\newblock {\em J. Ocean Eng. Mar. Energy}, 2(3):251--278, apr 2016.

\bibitem{Gotoh2021}
Hitoshi Gotoh, Abbas Khayyer, and Yuma Shimizu.
\newblock {\em Appl. Ocean Res.}, 115:102822, oct 2021.

\bibitem{Koshizuka1996}
S.~Koshizuka and Y.~Oka.
\newblock {\em Nucl. Sci. Eng.}, 123(3):421--434, 1996.

\bibitem{Gingold1977}
R.~A. Gingold and J.~J. Monaghan.
\newblock {\em Mon. Not. R. Astron. Soc.}, 181(3):375--389, dec 1977.

\bibitem{Monaghan1994}
J.~J. Monaghan.
\newblock {\em J. Comput. Phys.}, 110(2):399--406, feb 1994.

\bibitem{Koshizuka1998}
S.~Koshizuka, Atsushi Nobe, and Y.~Oka.
\newblock {\em Int. J. Numer. Methods Fluids}, 26(7):751--769, 1998.

\bibitem{Xu2009}
Rui Xu, Peter Stansby, and Dominique Laurence.
\newblock {\em J. Comput. Phys.}, 228(18):6703--6725, oct 2009.

\bibitem{Khayyer2011}
Abbas Khayyer and Hitoshi Gotoh.
\newblock {\em J. Comput. Phys.}, 230(8):3093--3118, apr 2011.

\bibitem{Tamai2014}
Tasuku Tamai and Seiichi Koshizuka.
\newblock {\em Comput. Part. Mech.}, 1(3):277--305, sep 2014.

\bibitem{Souto-Iglesias2013}
Antonio Souto-Iglesias, Fabricio MacI{\`{a}}, Leo~M. Gonz{\'{a}}lez, and
  Jose~L. Cercos-Pita.
\newblock {\em Comput. Phys. Commun.}, 184(3):732--745, mar 2013.

\bibitem{Duan2019}
Guangtao Duan, Akifumi Yamaji, Seiichi Koshizuka, and Bin Chen.
\newblock {\em Comput. Fluids}, 190:254--273, aug 2019.

\bibitem{Li2020}
Gen Li, Jinchen Gao, Panpan Wen, Quanbin Zhao, Jinshi Wang, Junjie Yan, and
  Akifumi Yamaji, aug 2020.

\bibitem{Jeffery1922}
G.~B. Jeffery.
\newblock {\em Proc. R. Soc. London, Ser. A}, 102(715):161--179, nov 1922.

\bibitem{Meyer2020}
Nils Meyer, Oleg Saburow, Martin Hohberg, Andrew~N. Hrymak, Frank Henning, and
  Luise K{\"{a}}rger.
\newblock {\em J. Compos. Sci.}, 4(2):77, jun 2020.

\bibitem{Yashiro2011}
S.~Yashiro, T.~Okabe, and K.~Matsushima.
\newblock {\em Adv. Compos. Mater.}, 20(6):503--517, 2011.

\bibitem{Yashiro2012}
S.~Yashiro, Hideaki Sasaki, and Yoshihisa Sakaida.
\newblock {\em Compos. Part A Appl. Sci. Manuf.}, 43(10):1754--1764, oct 2012.

\bibitem{Eringen1966a}
A.~Eringen.
\newblock {\em Indiana Univ. Math. J.}, 16(1):1--18, 1966.

\bibitem{Souto-Iglesias2021}
A.~Souto-Iglesias, J.~Bonet Avalos, M.~Antuono, and A.~Colagrossi.
\newblock {\em Phys. Rev. E}, 104(1):015315, jul 2021.

\bibitem{Oochi2010}
M.~Oochi, S.~Koshizuka, and M.~Sakai.
\newblock {\em Trans. Japan Soc. Comput. Eng. Sci.}, 2010:20100013--20100013,
  2010.

\bibitem{Shakibaeinia2010}
Ahmad Shakibaeinia and Yee~Chung Jin.
\newblock {\em Int. J. Numer. Methods Fluids}, 63(10):1208--1232, aug 2010.

\bibitem{Yamamoto1993}
Satoru Yamamoto and Takaaki Matsuoka.
\newblock {\em J. Chem. Phys.}, 98(1):644--650, 1993.

\bibitem{Kuzkin2012}
Vitaly~A. Kuzkin and Igor~E. Asonov.
\newblock {\em Phys. Rev. E}, 86(5):051301, nov 2012.

\bibitem{Colagrossi2012}
Andrea Colagrossi, B.~Bouscasse, M.~Antuono, and S.~Marrone.
\newblock {\em Comput. Phys. Commun.}, 183(8):1641--1653, aug 2012.

\bibitem{Einarsson2015}
J.~Einarsson, F.~Candelier, F.~Lundell, J.~R. Angilella, and B.~Mehlig.
\newblock {\em Phys. Fluids}, 27(6):063301, jun 2015.

\bibitem{Bretherton1962}
F.~P. Bretherton.
\newblock {\em J. Fluid Mech.}, 14(2):284--304, 1962.

\bibitem{Trevelyan1951}
B.~J. Trevelyan and S.~G. Mason.
\newblock {\em J. Colloid Sci.}, 6(4):354--367, aug 1951.

\bibitem{Saffman1956}
P.~G. Saffman.
\newblock {\em J. Fluid Mech.}, 1(5):540--553, 1956.

\bibitem{Fan1998}
Xijun Fan, N.~Phan-Thien, and Rong Zheng.
\newblock {\em J. Nonnewton. Fluid Mech.}, 74(1-3):113--135, jan 1998.

\bibitem{Yamamoto1994}
Satoru Yamamoto and Takaaki Matsuoka.
\newblock {\em J. Chem. Phys.}, 100(4):3317--3324, 1994.

\bibitem{Skjetne1997a}
Paal Skjetne, Russell~F. Ross, and Daniel~J. Klingenberg.
\newblock {\em J. Chem. Phys.}, 107(6):2108--2121, aug 1997.

\bibitem{Lindstrom2009}
Stefan~B. Lindstr{\"{o}}m and Tetsu Uesaka.
\newblock {\em Phys. Fluids}, 21(8):083301, aug 2009.

\bibitem{Sasayama2022}
Toshiki Sasayama, Hirotaka Okamoto, Norikazu Sato, and Jumpei Kawada.
\newblock {\em Powder Technol.}, 404:117481, may 2022.

\end{thebibliography}
\end{document}